\input harvmac
%

\let\includefigures=\iftrue
\let\useblackboard=\iftrue
\newfam\black

\includefigures
\message{If you do not have epsf.tex (to include figures),}
\message{change the option at the top of the tex file.}
\input epsf
\def\figin{\epsfcheck\figin}\def\figins{\epsfcheck\figins}
\def\epsfcheck{\ifx\epsfbox\UnDeFiNeD
\message{(NO epsf.tex, FIGURES WILL BE IGNORED)}
\gdef\figin##1{\vskip2in}\gdef\figins##1{\hskip.5in}
\else\message{(FIGURES WILL BE INCLUDED)}%
\gdef\figin##1{##1}\gdef\figins##1{##1}\fi}
\def\DefWarn#1{}
\def\figinsert{\goodbreak\midinsert}
\def\ifig#1#2#3{\DefWarn#1\xdef#1{fig.~\the\figno}
\writedef{#1\leftbracket fig.\noexpand~\the\figno}%
\figinsert\figin{\centerline{#3}}\medskip\centerline{\vbox{
\baselineskip12pt\advance\hsize by -1truein
\noindent\footnotefont{\bf Fig.~\the\figno:} #2}}
\endinsert\global\advance\figno by1}
\else
\def\ifig#1#2#3{\xdef#1{fig.~\the\figno}
\writedef{#1\leftbracket fig.\noexpand~\the\figno}%
\global\advance\figno by1} \fi

\def\journal#1&#2(#3){\unskip, \sl #1\ \bf #2 \rm(19#3) }
\def\andjournal#1&#2(#3){\sl #1~\bf #2 \rm (19#3) }

\def\ie{{\it i.e.}}
\def\eg{{\it e.g.}}

\noblackbox
%


\def\unlockat{\catcode`\@=11}
\def\lockat{\catcode`\@=12}

\unlockat

\def\newsec#1{\global\advance\secno by1\message{(\the\secno. #1)}
\global\subsecno=0\global\subsubsecno=0\eqnres@t\noindent
{\bf\the\secno. #1}
\writetoca{{\secsym} {#1}}\par\nobreak\medskip\nobreak}
\global\newcount\subsecno \global\subsecno=0
\def\subsec#1{\global\advance\subsecno
by1\message{(\secsym\the\subsecno. #1)}
\ifnum\lastpenalty>9000\else\bigbreak\fi\global\subsubsecno=0
\noindent{\it\secsym\the\subsecno. #1}
\writetoca{\string\quad {\secsym\the\subsecno.} {#1}}
\par\nobreak\medskip\nobreak}
\global\newcount\subsubsecno \global\subsubsecno=0
\def\subsubsec#1{\global\advance\subsubsecno by1
\message{(\secsym\the\subsecno.\the\subsubsecno. #1)}
\ifnum\lastpenalty>9000\else\bigbreak\fi
\noindent\quad{\secsym\the\subsecno.\the\subsubsecno.}{#1}
\writetoca{\string\qquad{\secsym\the\subsecno.\the\subsubsecno.}{#1}}
\par\nobreak\medskip\nobreak}

\def\subsubseclab#1{\DefWarn#1\xdef
#1{\noexpand\hyperref{}{subsubsection}%
{\secsym\the\subsecno.\the\subsubsecno}%
{\secsym\the\subsecno.\the\subsubsecno}}%
\writedef{#1\leftbracket#1}\wrlabeL{#1=#1}}
\lockat

\def\ie{{\it i.e.}}
\def\eg{{\it e.g.}}

\def\CM {{\cal M}}

\font\manual=manfnt \def\dbend{\lower3.5pt\hbox{\manual\char127}}

\def\IZ{\relax\ifmmode\mathchoice
{\hbox{\cmss Z\kern-.4em Z}}{\hbox{\cmss Z\kern-.4em Z}}
{\lower.9pt\hbox{\cmsss Z\kern-.4em Z}}
{\lower1.2pt\hbox{\cmsss Z\kern-.4em Z}}\else{\cmss Z\kern-.4em
Z}\fi}

\def\CM {{\cal M}}


\def\IZ{\relax\ifmmode\mathchoice
{\hbox{\cmss Z\kern-.4em Z}}{\hbox{\cmss Z\kern-.4em Z}}
{\lower.9pt\hbox{\cmsss Z\kern-.4em Z}}
{\lower1.2pt\hbox{\cmsss Z\kern-.4em Z}}\else{\cmss Z\kern-.4em
Z}\fi}
\def\IB{\relax{\rm I\kern-.18em B}}
\def\IC{{\relax\hbox{$\inbar\kern-.3em{\rm C}$}}}
\def\ID{\relax{\rm I\kern-.18em D}}
\def\IE{\relax{\rm I\kern-.18em E}}
\def\IF{\relax{\rm I\kern-.18em F}}
\def\IG{\relax\hbox{$\inbar\kern-.3em{\rm G}$}}
\def\IGa{\relax\hbox{${\rm I}\kern-.18em\Gamma$}}
\def\IH{\relax{\rm I\kern-.18em H}}
\def\II{\relax{\rm I\kern-.18em I}}
\def\IK{\relax{\rm I\kern-.18em K}}
\def\IP{\relax{\rm I\kern-.18em P}}
\def\IQ{\relax\hbox{$\inbar\kern-.3em{\rm Q}$}}

\def\inbar{\,\vrule height1.5ex width.4pt depth0pt}

\font\cmss=cmss10 \font\cmsss=cmss10 at 7pt
\def\IR{\relax{\rm I\kern-.18em R}}

%
%

\def\makeblankbox#1#2{\hbox{\lower\dp0\vbox{\hidehrule{#1}{#2}%
   \kern -#1
   \hbox to \wd0{\hidevrule{#1}{#2}%
      \raise\ht0\vbox to #1{}
      \lower\dp0\vtop to #1{}
      \hfil\hidevrule{#2}{#1}}%
   \kern-#1\hidehrule{#2}{#1}}}%
}%
\def\hidehrule#1#2{\kern-#1\hrule height#1 depth#2 \kern-#2}%
\def\hidevrule#1#2{\kern-#1{\dimen0=#1\advance\dimen0 by #2\vrule
    width\dimen0}\kern-#2}%
\def\openbox{\ht0=1.2mm \dp0=1.2mm \wd0=2.4mm  \raise 2.75pt
\makeblankbox {.25pt} {.25pt}  }

\def\bun#1/#2{\leavevmode
   \kern.1em \raise .5ex \hbox{\the\scriptfont0 #1}%
   \kern-.1em $/$%
   \kern-.15em \lower .25ex \hbox{\the\scriptfont0 #2}%
}

\def\opensquare{\ht0=3.4mm \dp0=3.4mm \wd0=6.8mm  \raise 2.7pt
\makeblankbox {.25pt} {.25pt}  }


\def\sector#1#2{\ {\scriptstyle #1}\hskip 1mm
\mathop{\opensquare}\limits_{\lower 1mm\hbox{$\scriptstyle#2$}}\hskip 1mm}

\def\tsector#1#2{\ {\scriptstyle #1}\hskip 1mm
\mathop{\opensquare}\limits_{\lower 1mm\hbox{$\scriptstyle#2$}}^\sim\hskip 1mm}


\def\inbar{\,\vrule height1.5ex width.4pt depth0pt}

\font\cmss=cmss10 \font\cmsss=cmss10 at 7pt
\def\IR{\relax{\rm I\kern-.18em R}}


\def\frac#1#2{{#1\over#2}}

\def\inbar{\,\vrule height1.5ex width.4pt depth0pt}
\def\IC{\relax\hbox{$\inbar\kern-.3em{\rm C}$}}
\def\IR{\relax{\rm I\kern-.18em R}}
\def\IP{\relax{\rm I\kern-.18em P}}

%
%
\catcode`\@=11
\def\slash#1{\mathord{\mathpalette\c@ncel{#1}}}
\overfullrule=0pt

\def\II{{\cal I}}

\def\underrel#1\over#2{\mathrel{\mathop{\kern\z@#1}\limits_{#2}}}

\catcode`\@=12


%



\def\frac#1#2{{#1\over#2}}

\def\inbar{\,\vrule height1.5ex width.4pt depth0pt}
\def\IC{\relax\hbox{$\inbar\kern-.3em{\rm C}$}}
\def\IR{\relax{\rm I\kern-.18em R}}
\def\IP{\relax{\rm I\kern-.18em P}}

%
%

%
\catcode`\@=11
\def\slash#1{\mathord{\mathpalette\c@ncel{#1}}}
\overfullrule=0pt

\def\II{{\cal I}}

\def\underrel#1\over#2{\mathrel{\mathop{\kern\z@#1}\limits_{#2}}}

\catcode`\@=12


%


\lref\GaiottoQI{
  D.~Gaiotto and X.~Yin,
  ``Notes on superconformal Chern-Simons-matter theories,''
  JHEP {\bf 0708}, 056 (2007)
  [arXiv:0704.3740 [hep-th]].
}

\lref\SchwarzYJ{
  J.~H.~Schwarz,
  ``Superconformal Chern-Simons theories,''
  JHEP {\bf 0411}, 078 (2004)
  [arXiv:hep-th/0411077].
}

\lref\WittenDS{
  E.~Witten,
  ``Supersymmetric index of three-dimensional gauge theory,''
  arXiv:hep-th/9903005.
}

\lref\HananyIE{
  A.~Hanany and E.~Witten,
  ``Type IIB superstrings, BPS monopoles, and three-dimensional gauge
  dynamics,''
  Nucl.\ Phys.\  B {\bf 492}, 152 (1997)
  [arXiv:hep-th/9611230].
}

\lref\GiveonSR{
  A.~Giveon and D.~Kutasov,
  ``Brane dynamics and gauge theory,''
  Rev.\ Mod.\ Phys.\  {\bf 71}, 983 (1999)
  [arXiv:hep-th/9802067].
}

\lref\AharonyJU{
  O.~Aharony and A.~Hanany,
  ``Branes, superpotentials and superconformal fixed points,''
  Nucl.\ Phys.\  B {\bf 504}, 239 (1997)
  [arXiv:hep-th/9704170].
}

\lref\ElitzurFH{
  S.~Elitzur, A.~Giveon and D.~Kutasov,
  ``Branes and N = 1 duality in string theory,''
  Phys.\ Lett.\  B {\bf 400}, 269 (1997)
  [arXiv:hep-th/9702014].
}

\lref\BergmanNA{
  O.~Bergman, A.~Hanany, A.~Karch and B.~Kol,
  ``Branes and supersymmetry breaking in 3D gauge theories,''
  JHEP {\bf 9910}, 036 (1999)
  [arXiv:hep-th/9908075].
}

\lref\OhtaIV{
  K.~Ohta,
  ``Supersymmetric index and s-rule for type IIB branes,''
  JHEP {\bf 9910}, 006 (1999)
  [arXiv:hep-th/9908120].
}

\lref\KitaoMF{
  T.~Kitao, K.~Ohta and N.~Ohta,
  ``Three-dimensional gauge dynamics from brane configurations with
  (p,q)-fivebrane,''
  Nucl.\ Phys.\  B {\bf 539}, 79 (1999)
  [arXiv:hep-th/9808111].
}

\lref\GiveonWP{
  A.~Giveon, A.~Katz and Z.~Komargodski,
  ``On SQCD with massive and massless flavors,''
  JHEP {\bf 0806}, 003 (2008)
  [arXiv:0804.1805 [hep-th]].
}

\lref\BrodieSZ{
  J.~H.~Brodie and A.~Hanany,
  ``Type IIA superstrings, chiral symmetry, and N = 1 4D gauge theory
  dualities,''
  Nucl.\ Phys.\  B {\bf 506}, 157 (1997)
  [arXiv:hep-th/9704043].
}

\lref\AharonyGK{
  O.~Aharony, O.~Bergman and D.~L.~Jafferis,
  ``Fractional M2-branes,''
  arXiv:0807.4924 [hep-th].
}

\lref\NovikovUC{
  V.~A.~Novikov, M.~A.~Shifman, A.~I.~Vainshtein and V.~I.~Zakharov,
  ``Exact Gell-Mann-Low Function Of Supersymmetric Yang-Mills Theories From
  Instanton Calculus,''
  Nucl.\ Phys.\  B {\bf 229}, 381 (1983).
}

\lref\SeibergPQ{
  N.~Seiberg,
  ``Electric - magnetic duality in supersymmetric nonAbelian gauge theories,''
  Nucl.\ Phys.\  B {\bf 435}, 129 (1995)
  [arXiv:hep-th/9411149].
}

\lref\IntriligatorJJ{
  K.~Intriligator and B.~Wecht,
  ``The exact superconformal R-symmetry maximizes a,''
  Nucl.\ Phys.\  B {\bf 667}, 183 (2003)
  [arXiv:hep-th/0304128].
}

\lref\KutasovIY{
  D.~Kutasov, A.~Parnachev and D.~A.~Sahakyan,
  ``Central charges and U(1)R symmetries in N = 1 super Yang-Mills,''
  JHEP {\bf 0311}, 013 (2003)
  [arXiv:hep-th/0308071].
}

\lref\IntriligatorMI{
  K.~Intriligator and B.~Wecht,
  ``RG fixed points and flows in SQCD with adjoints,''
  Nucl.\ Phys.\  B {\bf 677}, 223 (2004)
  [arXiv:hep-th/0309201].
}

\lref\KutasovUX{
  D.~Kutasov,
  ``New results on the 'a-theorem' in four dimensional supersymmetric field
  theory,''
  arXiv:hep-th/0312098.
}

\lref\IntriligatorDD{
  K.~Intriligator, N.~Seiberg and D.~Shih,
  ``Dynamical SUSY breaking in meta-stable vacua,''
  JHEP {\bf 0604}, 021 (2006)
  [arXiv:hep-th/0602239].
}

\lref\OoguriBG{
  H.~Ooguri and Y.~Ookouchi,
  ``Meta-stable supersymmetry breaking vacua on intersecting branes,''
  Phys.\ Lett.\ B {\bf 641}, 323 (2006)
  [arXiv:hep-th/0607183].
}

\lref\FrancoHT{
  S.~Franco, I.~Garcia-Etxebarria and A.~M.~Uranga,
  ``Non-supersymmetric meta-stable vacua from brane configurations,''
  JHEP {\bf 0701}, 085 (2007)
  [arXiv:hep-th/0607218].
}

\lref\BenaRG{
  I.~Bena, E.~Gorbatov, S.~Hellerman, N.~Seiberg and D.~Shih,
  ``A note on (meta)stable brane configurations in MQCD,''
  JHEP {\bf 0611}, 088 (2006)
  [arXiv:hep-th/0608157].
}

\lref\GiveonFK{
  A.~Giveon and D.~Kutasov,
  ``Gauge symmetry and supersymmetry breaking from intersecting branes,''
  Nucl.\ Phys.\  B {\bf 778}, 129 (2007)
  [arXiv:hep-th/0703135].
}

\lref\GiveonEF{
  A.~Giveon and D.~Kutasov,
  ``Stable and Metastable Vacua in SQCD,''
  Nucl.\ Phys.\  B {\bf 796}, 25 (2008)
  [arXiv:0710.0894 [hep-th]].
}

\lref\GiveonEW{
  A.~Giveon and D.~Kutasov,
  ``Stable and Metastable Vacua in Brane Constructions of SQCD,''
  JHEP {\bf 0802}, 038 (2008)
  [arXiv:0710.1833 [hep-th]].
}

\lref\AharonyUG{
  O.~Aharony, O.~Bergman, D.~L.~Jafferis and J.~Maldacena,
  ``N=6 superconformal Chern-Simons-matter theories, M2-branes and their
  gravity duals,''
  arXiv:0806.1218 [hep-th].
}

\lref\KutasovVE{
  D.~Kutasov,
  ``A Comment on duality in N=1 supersymmetric nonAbelian gauge theories,''
  Phys.\ Lett.\  B {\bf 351}, 230 (1995)
  [arXiv:hep-th/9503086].
}

\lref\KutasovNP{
  D.~Kutasov and A.~Schwimmer,
  ``On duality in supersymmetric Yang-Mills theory,''
  Phys.\ Lett.\  B {\bf 354}, 315 (1995)
  [arXiv:hep-th/9505004].
}

\lref\KutasovSS{
  D.~Kutasov, A.~Schwimmer and N.~Seiberg,
  ``Chiral Rings, Singularity Theory and Electric-Magnetic Duality,''
  Nucl.\ Phys.\  B {\bf 459}, 455 (1996)
  [arXiv:hep-th/9510222].
}

\lref\KarchUX{
  A.~Karch,
  ``Seiberg duality in three dimensions,''
  Phys.\ Lett.\  B {\bf 405}, 79 (1997)
  [arXiv:hep-th/9703172].
}

\lref\AharonyGP{
  O.~Aharony,
  ``IR duality in d = 3 N = 2 supersymmetric USp(2N(c)) and U(N(c)) gauge
  theories,''
  Phys.\ Lett.\  B {\bf 404}, 71 (1997)
  [arXiv:hep-th/9703215].
}

\lref\KachruAW{
  S.~Kachru, R.~Kallosh, A.~Linde and S.~P.~Trivedi,
  ``De Sitter vacua in string theory,''
  Phys.\ Rev.\  D {\bf 68}, 046005 (2003)
  [arXiv:hep-th/0301240].
}

\Title{
} {\vbox{ \centerline{Seiberg Duality in Chern-Simons Theory}}}
\medskip
\centerline{\it Amit Giveon${}^{1}$ and David Kutasov${}^{2}$}
\bigskip
\smallskip
\centerline{${}^{1}$Racah Institute of Physics, The Hebrew
University} \centerline{Jerusalem 91904, Israel}
\smallskip
\centerline{${}^2$EFI and Department of Physics, University of
Chicago} \centerline{5640 S. Ellis Av., Chicago, IL 60637, USA }

\bigskip\bigskip\bigskip
\noindent

We argue that $N=2$ supersymmetric Chern-Simons  theories exhibit
a strong-weak coupling Seiberg-type duality. We also discuss
supersymmetry breaking in these theories.

\vglue .3cm
\bigskip

\Date{8/08}

\bigskip

\newsec{Introduction}

Three dimensional Chern-Simons (CS) gauge theories with $N=2$ supersymmetry
(\ie\ four real supercharges) coupled to ``matter'' chiral superfields give
rise to a large class of quantum field theories with non-trivial infrared
dynamics (see \eg\ \refs{\SchwarzYJ,\GaiottoQI} and references therein).
These theories are characterized by a gauge group $G$, Chern-Simons level $k$,
and matter representation $R$. They are classically
conformal, since the level $k$, which plays the role of a coupling constant,
is dimensionless. The conformal symmetry extends to the quantum theory since
$k$ does not run along Renormalization Group (RG) trajectories.
Indeed, for non-Abelian gauge groups $k$ is quantized, and thus cannot run.

One can also add superpotential interactions among the matter superfields.
In general, these break the conformal symmetry and generate non-trivial RG
flows. In some cases they modify the infrared behavior.

Determining the quantum dynamics of these theories is an interesting problem,
which in many ways is reminiscent of the analogous problem in four dimensional
Yang-Mills theories with $N=1$ supersymmetry. However, while in four dimensions
much progress has been made by using the NSVZ $\beta$-function \NovikovUC,
Seiberg duality \SeibergPQ, $a$-maximization \refs{\IntriligatorJJ\KutasovIY
\IntriligatorMI-\KutasovUX}, etc, in three dimensional CS theory the understanding
is more rudimentary. For large $k$ one can use perturbation theory in $1/k$,
but in general the problem is unsolved.

As mentioned above, one of the important tools in analyzing the infrared
dynamics of four dimensional $N=1$ supersymmetric gauge theories is Seiberg
duality, which in many cases maps a strongly coupled gauge theory to a weakly
coupled or IR free one. In this note we will propose an analog of Seiberg
duality for three dimensional $N=2$ supersymmetric Chern-Simons theories.
While this duality is a field theory phenomenon, we will phrase the discussion
in terms of brane constructions that reduce to the relevant field theories at
low energies. These constructions capture efficiently both classical and
quantum aspects of CS dynamics.

The rest of this note is organized as follows. We start by describing the brane
configuration we will be interested in and its low energy CS description. We
then use the results of \ElitzurFH\ to construct the Seiberg dual configuration
and analyze its low energy limit. We discuss the relation between the two
CS theories, and propose that they are equivalent. We also describe supersymmetry
breaking vacua that generalize the ISS \IntriligatorDD\ construction to CS
theory. Unlike their four dimensional analogs, these vacua appear to be stable.

\newsec{Electric theory}


We will study brane configurations in type IIB string theory that
involve two types of $NS5$-branes, which we will denote by $NS$
and $NS'$, as well as $D3$-branes and $D5$-branes. The different
branes are oriented as follows in $\IR^{9,1}$:
\eqn\branes{\eqalign{ NS:&\qquad(012345)\cr NS':&\qquad(012389)\cr
D3:&\qquad(0126)\cr D5:&\qquad(012789)\cr }} These are precisely
the branes that are used in the Hanany-Witten construction
\HananyIE\ of three dimensional gauge theories (see \GiveonSR\ for
a review). It is not difficult to check that a configuration which
includes all the branes in \branes\ preserves $N=2$ supersymmetry
in the three dimensions common to all the branes, $(012)$.

\ifig\loc{Fivebrane recombination.}
{\epsfxsize4.8in\epsfbox{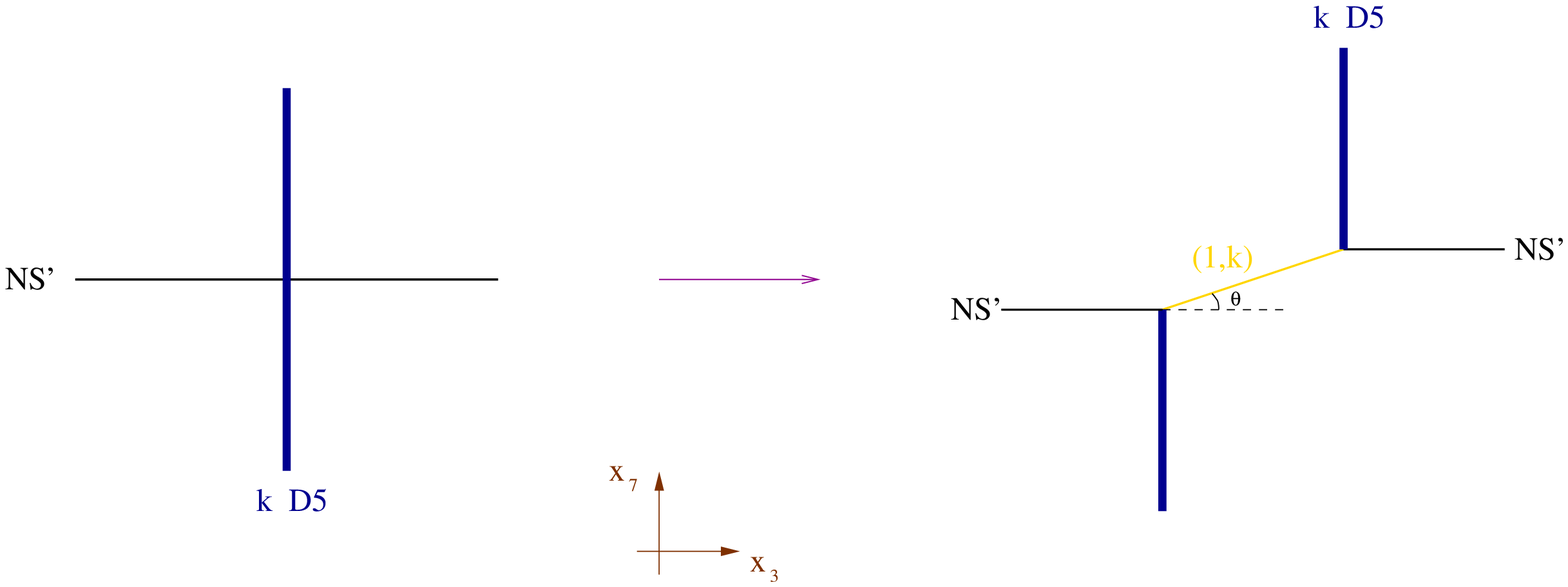}}

When an $NS'$-brane intersects $k$ $D5$-branes in the
$(37)$-plane, the two types of branes can locally combine into a
$(1,k)$ fivebrane (see figure 1), which is oriented at an angle
$\theta$ to the $NS'$-brane, with $\tan\theta=g_sk$ \AharonyJU.
The resulting brane configuration preserves supersymmetry for all
values of the length of the $(1,k)$ fivebrane segment. When the
length of this segment goes to infinity, the $NS'$-brane and
$D5$-branes are replaced by the $(1,k)$ fivebrane everywhere; the
supersymmetry is not affected.


The brane configuration we consider is depicted in figure 2a,
where we use the notation: \eqn\vwy{v=x^4+ix^5,\qquad
w=x^8+ix^9,\qquad y=x^6~.} The corresponding low energy theory is a
$U(N_c)$ gauge theory with $N_f+k$ flavors of chiral superfields
$Q^i$, $\tilde Q_i$ in the fundamental representation of the gauge
group \GiveonSR.

\ifig\loc{Electric brane configuration.}
{\epsfxsize5.0in\epsfbox{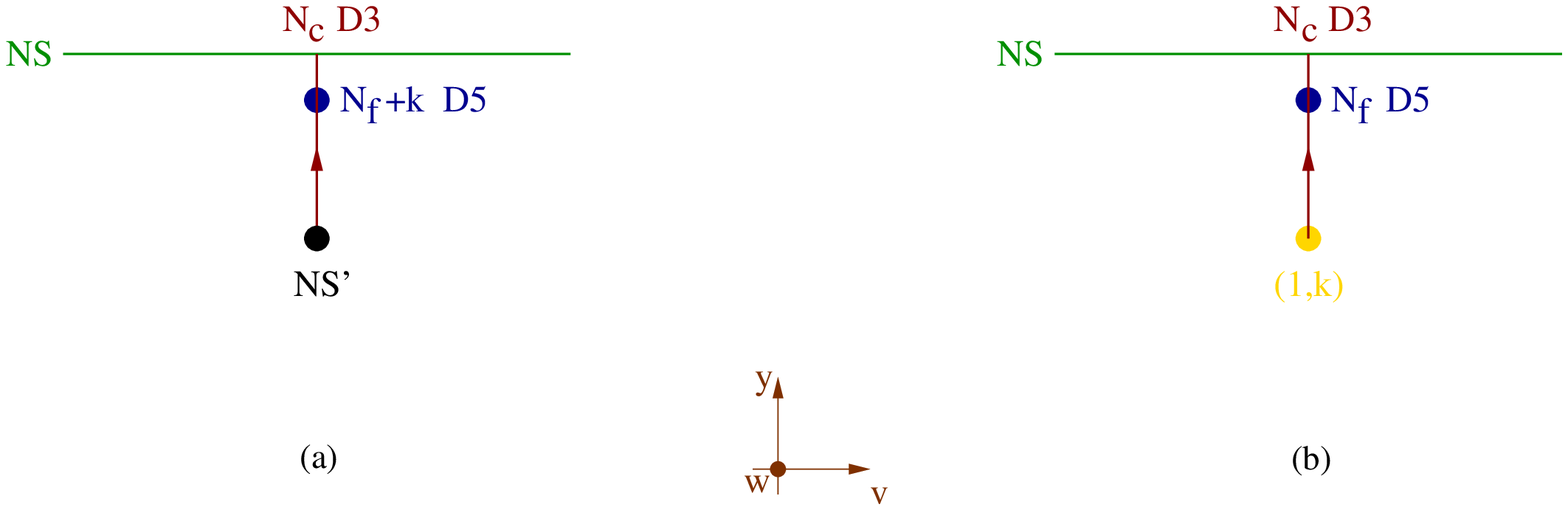}}

\noindent
In order to study the dynamics of interest, we move $k$ of the
$D5$-branes towards the $NS'$-brane, and when the two intersect,
deform the configuration as in figure 1, such that the $NS'$-brane
and $k$ $D5$-branes are replaced by a $(1,k)$ fivebrane. The
resulting brane configuration appears in figure 2b. The
deformation that takes figure 2a to 2b corresponds in the field
theory to giving real masses of the same sign to the $k$ flavors
of fundamentals $Q$ and $\tilde Q$ that were singled out in the
construction \BergmanNA. The limit in figure 2b corresponds to
sending these masses to infinity.

The low energy limit of the brane configuration of figure 2b is
described by a level $k$ $U(N_c)$ CS theory \KitaoMF, coupled to
$N_f$ fundamentals $Q^i$, $\tilde Q_i$, $i=1,2,\cdots, N_f$. It
preserves $N=2$ superconformal symmetry. In the remainder of this
section we briefly comment on some of its properties.

The global symmetry of the gauge theory is $SU(N_f)\times
SU(N_f)\times U(1)_a\times U(1)_R$. The first three factors can be
seen in the brane picture by starting with the configuration of
figure 2b, moving all $N_f$ $D5$-branes to the $(1,k)$-brane, and
performing separate $U(N_f)$ transformations on the portions of
the $D5$-branes with $x^7>0$ and $x^7<0$ \BrodieSZ. The $U(1)_R$
is a subgroup of the $9+1$ dimensional Lorentz group preserved by
the brane configuration.

One can also use the brane picture to identify some of the
perturbations and moduli of the low energy field theory \GiveonSR.
In particular, moving the $D5$-branes in the $v$ direction
corresponds to turning on complex masses for $Q^i$, $\tilde Q_i$
via a superpotential of the form $W=m_i\tilde Q_i Q^i$. Moving
them in the $x^3$ direction corresponds to giving real masses with
opposite signs to $Q$, $\tilde Q$.

The moduli of the CS theory can be seen geometrically exactly as
in the four dimensional $N=1$ case, by separating the $N_f$
$D5$-branes in figure 2b in $x^6$, and allowing the $D3$-branes to
break on them (see \eg\ figure 25 in \GiveonSR). One finds, as
there, that the dimension of the moduli space is given by
\eqn\dimhiggs{{\rm dim}\CM=\cases{N_f^2&$N_f<N_c$\cr
2N_cN_f-N_c^2& $N_f\ge N_c$}}

\noindent
The classical analysis above receives quantum corrections due to
the following effect. It was shown in \refs{\BergmanNA,\OhtaIV} that 
the number of $D3$-branes that can stretch between an $NS$-brane and 
a $(1,k)$-brane without breaking supersymmetry is bounded from above 
by $k$. This is a consequence of the ``s-rule'' of \HananyIE, and 
is related to the fact that such $D3$-branes are necessarily on top 
of each other. At first sight it seems that this implies that in the 
configuration of figure 2b there is no supersymmetric vacuum unless 
$N_c\le k$, but the actual bound is less restrictive.

The reason is that one can think of $N_f$ out of the $N_c$
$D3$-branes\foot{It is enough to consider the case $N_c\ge N_f$.}
in figure 2b as stretching from the $NS$-brane to the $D5$-branes
and then from the $D5$-branes to the $(1,k)$-brane, so the net
number of $D3$-branes that enters the bound of \refs{\BergmanNA,\OhtaIV} 
is $N_c-N_f$. Hence, we conclude that the CS theory corresponding
to figure 2b has a supersymmetric vacuum for
\eqn\vaccond{N_f+k-N_c\ge 0~.} When \vaccond\ is satisfied, the
quantum moduli space has the dimension \dimhiggs. Note that the
constraint \vaccond\ allows $N_f$ to be either smaller or larger
than $N_c$. Note also that although we presented the derivation of
\vaccond\ in brane terms, it is a property of the CS theory
\refs{\WittenDS,\BergmanNA,\OhtaIV}.

\newsec{Magnetic theory and duality}


In order to construct the dual theory, we follow \ElitzurFH\ and
exchange the $NS$ and $(1,k)$ fivebranes. A convenient way to do
this is to go back to the configuration of figure 2a, move all
$N_f+k$ $D5$-branes to the other side of the $NS$-brane, creating
$N_f+k$ $D3$-branes in the process \HananyIE, and then move the
$NS'$-brane through the $NS$-brane. Finally, we
need to recombine the $k$ $D5$-branes with the $NS'$-brane into a
$(1,k)$-brane, as in the transition from figure 2a to 2b. The
resulting brane configuration is depicted in figure 3.

\ifig\loc{Magnetic brane configuration.}
{\epsfxsize2.4in\epsfbox{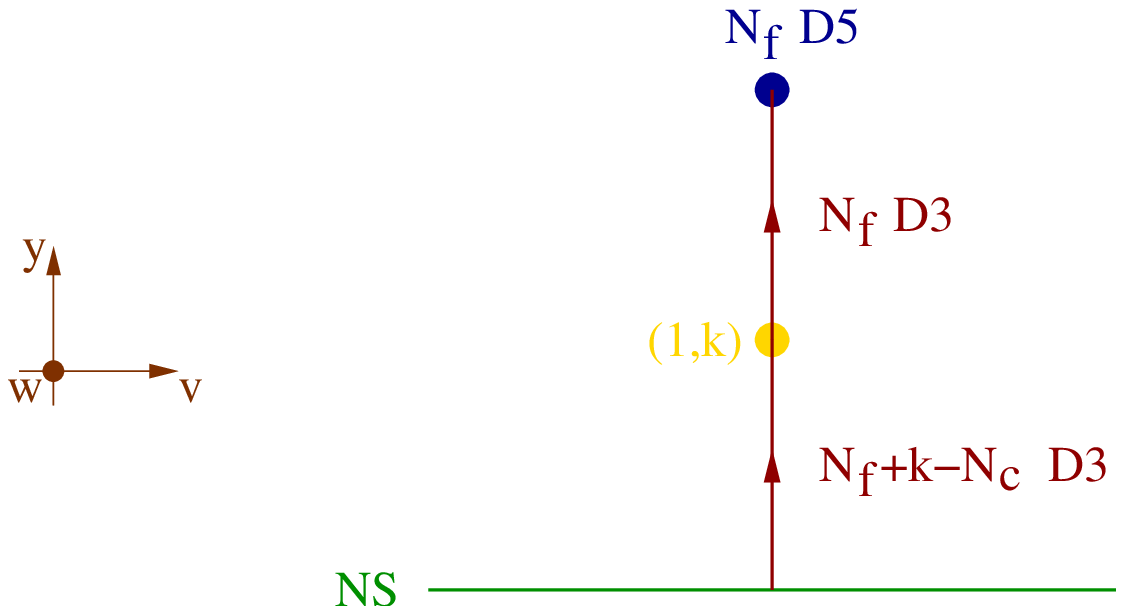}}

The low energy effective field theory can be read off figure 3 as
in \refs{\GiveonSR,\KitaoMF}. It includes a level $k$
$U(N_f+k-N_c)$ CS gauge field coupled to $N_f$ fundamentals $q_i$,
$\tilde q^i$, as well as an $N_f\times N_f$ matrix of singlets
$M^i_j$, which couple to the fundamentals via the superpotential
\eqn\magsup{W=M^i_jq_i\tilde q^j~.} It is thus natural to propose
that this magnetic CS theory is dual to the electric one discussed in
the previous section, with the usual identification
\eqn\meson{M^i_j=Q^i\tilde Q_j~.} Note that the constraint \vaccond,
which is necessary for having a supersymmetric vacuum, is in the
magnetic theory just the requirement that the rank of the magnetic
gauge group is non-negative. This is reminiscent of what happens
in four dimensional $N=1$ supersymmetric QCD, where the analogous
constraint is $N_f-N_c\ge 0$.

Note also that unlike the four dimensional case, it is important here
that the duality involves $U(N_c)$ and $U(N_f+k-N_c)$ and not the corresponding
$SU$ groups. Indeed, the $U(1)$ factor is interacting in this case, and it
is easy to see that if it was not gauged, the duality could not be correct.

As a check of the duality, we may ask whether the magnetic CS
theory reproduces the moduli space of vacua of the electric
theory, whose dimension is given by \dimhiggs. Naively, it looks
like the moduli space of the brane configuration of figure 3 is
$N_f^2$ dimensional, with the counting being the same as in
figure 29 in \GiveonSR.

For $N_f\le N_c$ this answer is correct, but for $N_f>N_c$ it is important to
take into account the constraint on the number of $D3$-branes stretched between
the $NS$ and $(1,k)$ fivebranes, which played a role in the derivation of \vaccond.
Indeed, in this case, at a generic point in the $N_f^2$ dimensional classical
moduli space of figure 29 of \GiveonSR, we have in figure 3 $N_f+k-N_c>k$
$D3$-branes stretched between the fivebranes, which as mentioned before leads to a
non-supersymmetric state. To preserve supersymmetry, we must keep $N_f-N_c$ of the
flavor $D3$-branes at the origin. It is easy to check that taking this into account
leads to precise agreement with the electric result \dimhiggs.

We see that while the constraint \vaccond\ arises from quantum effects in the
electric theory and is a classical property of the magnetic one, the opposite
happens in the analysis of the moduli space: the dimension \dimhiggs\ is obtained
classically in the electric theory, and requires quantum effects in the magnetic one.

Another class of deformations involves giving masses to some of the flavors.
Turning on the superpotential $W=m_1\tilde Q_1 Q^1$ in the electric theory,
corresponds in figure 2b to separating one of the $N_f$ $D5$-branes in the $v$
direction from the $D3$-branes. Integrating out $Q^1$, $\tilde Q_1$ amounts
to sending this separation to infinity. In the magnetic configuration, of figure 3,
this deformation requires the $D3$-brane connected to that $D5$-brane to combine
with one of the $N_f+k-N_c$ $D3$-branes stretched between the $(1,k)$ and $NS$
fivebranes, thus reducing the rank of the gauge group by 1. This leads, as in the
four dimensional case \refs{\SeibergPQ,\ElitzurFH}, to a dual pair with
$N_f\to N_f-1$, with all other parameters remaining the same.

Giving equal and opposite sign real masses to $Q^1$, $\tilde Q_1$ corresponds in
figure 2b to moving the corresponding $D5$-brane away from the $D3$-branes in the
$x^3$ direction. There are now two types of supersymmetric vacua. In one, the
electric gauge group remains unbroken, \ie\ the $D3$-branes continue to stretch
between the $NS$ and $(1,k)$ fivebranes. Sending the displaced $D5$-brane to infinity
amounts to reducing $N_f$ by one unit while keeping all the other parameters fixed,
as before.

A second vacuum is obtained by allowing one of the $N_c$ $D3$-branes to break on
the displaced $D5$-brane, such that as it moves in $x^3$, half of the $D3$-brane
stretches between the $NS$ and $D5$ branes, while the other half stretches between
the $D5$ and $(1,k)$ branes. As the $D5$-brane is sent to infinity, one finds a
vacuum of the original kind, with both $N_f$ and $N_c$ reduced by one unit.

In the magnetic brane configuration of figure 3, one finds the same vacua. Displacing
one of the $D5$-branes in $x^3$, one finds again two types of supersymmetric configurations.
In one, the $D5$-brane drags with it the $D3$-brane attached to it, reducing $N_f$
by one but not changing the rank of the magnetic gauge group. This gives rise to the
magnetic dual of the second kind of electric vacuum discussed above.

The dual of the first kind of electric vacuum is obtained by reconnecting the $D3$-brane
attached to the mobile $D5$-brane to one of the color $D3$-branes, and then moving the
$D5$-brane in $x^3$. This gives rise to a vacuum in which both the number of flavor
and that of colors in the magnetic theory are reduced by one unit, in agreement with
expectations.

Finally, giving same sign real masses to $Q^1$, $\tilde Q_1$ corresponds
in the electric brane configuration of figure 2b to moving a $D5$-brane in $x^6$
towards the $(1,k)$ brane, and using the process of figure 1 to turn it into a
$(1,k+1)$ brane. This leads to the same type of theory, with $N_f\to N_f-1$,
$k\to k+1$.

Similarly, in the magnetic configuration of figure 3, we need to send a $D5$-brane
towards the $(1,k)$ fivebrane and make the transition of figure 1. This again
corresponds to taking $N_f\to N_f-1$ and $k\to k+1$. Note that the rank of the
magnetic gauge group does not change in the process, in agreement with the duality.

To summarize, we see that the duality proposed above is consistent with the structure
of moduli space and deformations. This duality is a strong-weak coupling
one in the following sense. Consider first the electric theory. The interactions
between the chiral superfields $Q^i$, $\tilde Q_i$ are due to the CS coupling $k$. Thus,
if we keep $N_c$, $N_f$ fixed and send $k\to\infty$, the electric theory becomes more and
more weakly coupled. Note that in this limit the quantum constraint \vaccond\ is automatically
satisfied, as one would expect. On the other hand, for $k$ of order $N_c$ the electric theory
is strongly coupled.

In the magnetic theory, we have two kinds of interactions. One is due to the $U(N_f+k-N_c)$
CS gauge field; the other due to the cubic superpotential \magsup. Let us first ignore the
superpotential and focus on the gauge interaction. In the regime where the electric CS
interaction is weakly coupled, the rank of the magnetic gauge group $\bar N_c=N_f+k-N_c$
is of order $k$. Thus, it is strongly coupled. To make the magnetic CS theory weakly coupled,
one needs to consider the regime $k\gg \bar N_c$. This can be achieved, for example, by keeping
$N_f$ and $\bar N_c$ fixed and sending $k\to\infty$. In this limit $N_c\simeq k$ so the electric
CS theory is strongly coupled.

Even when the magnetic CS gauge interaction is weak, the theory still contains a cubic
superpotential, \magsup, which is a relevant perturbation that grows in the infrared. We are
not going to say much about it here, except to note that:
\item{(1)} One can go to the regime $k\gg \bar N_c\gg 1$ with $N_f$ fixed (say), in which the
Wess-Zumino model with superpotential \magsup\ is a large $N$ vector model, which can presumably
be solved using standard large $N$ techniques. In this sense it is weakly coupled, with the small
coupling being $1/\bar N_c$.
\item{(2)} One can put the electric and magnetic theories on the same footing by adding to the
electric theory a quartic superpotential\foot{Such superpotentials in four dimensional $N=1$ SQCD
and their brane realizations have been recently studied in \refs{\GiveonEF,\GiveonEW}.}
\eqn\www{W=\lambda(\tilde QQ)^2~.}
Under the duality we proposed here, this corresponds to adding $W=\lambda M^2$ to \magsup.
Integrating out $M$ leads to a quartic superpotential for the magnetic quarks very similar to \www.
The resulting infrared theory preserves $N=3$ superconformal symmetry, and can be made arbitrarily
weakly coupled by tuning $k$, $N_f$ and $N_c$.

\newsec{Supersymmetry breaking}

In four dimensions, it was shown in \IntriligatorDD\ that the magnetic dual of  $N=1$ supersymmetric QCD
with a small mass deformation\foot{We restrict here to the case of equal masses for all the flavors.
In four dimensions, new effects appear when some of the masses are zero \GiveonWP; it would be interesting
to investigate the analogous problem in the CS case.},
\eqn\massdef{W=m\tilde Q_i Q^i~,}
has a metastable supersymmetry breaking vacuum. It is interesting to ask what happens in our case.
Consider first the electric theory. As discussed above, turning on the mass term \massdef\ corresponds
in the brane construction of figure 2b to displacing all $N_f$ $D5$-branes in the $v$ direction, by
the same amount. The resulting configuration has $N_c$ $D3$-branes stretched between the $NS$ and $(1,k)$
fivebranes with no $D5$-branes to screen them, so the s-rule implies that it is only supersymmetric when
\eqn\nncckk{N_c\le k~,}
a stronger constraint than \vaccond. In particular, for $N_c>k$ supersymmetry is spontaneously broken.

To connect to the discussion of \IntriligatorDD\ consider the magnetic theory of figure 3.
The mass deformation \massdef\ corresponds to adding to the magnetic superpotential \magsup\
the term $\delta W=mM$. In the brane construction, this corresponds again to moving the
$N_f$ $D5$-branes in the $v$ direction. This gives rise to the configuration of figure 4a.
This configuration is non-supersymmetric; its fate depends on whether the inequality \nncckk\
is satisfied. If it is, there are more color threebranes than flavor ones, so they reconnect
and lead to the configuration of figure 4b, which is the supersymmetric vacuum dual to that
discussed above in the electric theory.

\ifig\loc{Supersymmetric and non-supersymmetric vacua of the magnetic theory
for non-zero mass.}
{\epsfxsize5in\epsfbox{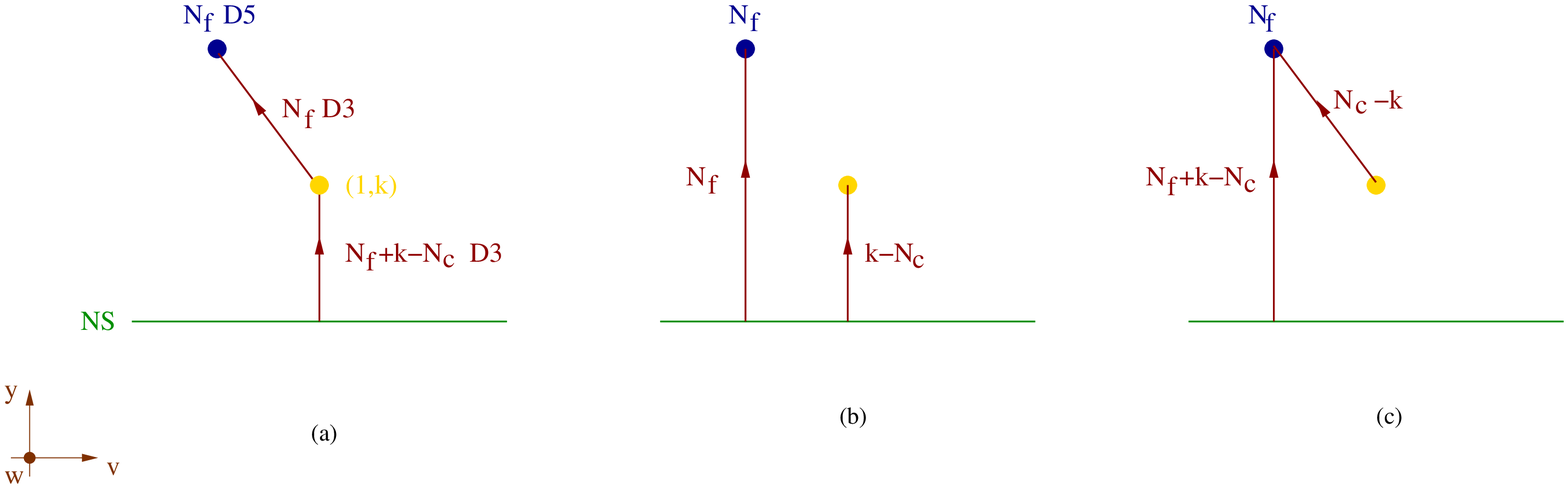}}

For $N_c>k$ there are not enough color branes to combine with all the flavor ones, and
the ground state of the system corresponds to the configuration of figure 4c. This
configuration is non-supersymmetric. It is clearly a direct analog of the four dimensional
configurations of \refs{\OoguriBG\FrancoHT\BenaRG-\GiveonFK}.
The difference is that while there these configurations were
metastable, and there was a supersysmmetric vacuum elsewhere in field space, here we expect
this supersymmetry breaking vacuum to be stable.

A quick way to see this is that in \refs{\OoguriBG\FrancoHT\BenaRG-\GiveonFK}
the electric brane configuration had supersymmetric
vacua, so the magnetic one must have them as well, by duality, whereas here the electric theory
breaks supersymmetry (for $N_c>k$). Also, in the four dimensional brane construction it is known
that certain quantum effects, which are needed for constructing the supersymmetric vacuum in the
magnetic theory, are difficult to see in the brane construction \GiveonSR, whereas in the three
dimensional brane constructions discussed here the quantum effects are expected to be visible in
the brane description.

Coming back to figure 4c, like in the four dimensional brane configurations of
\refs{\OoguriBG\FrancoHT\BenaRG-\GiveonFK}, the $N_c-k$ $D3$-branes stretched between the
$D5$-branes and the $(1,k)$ fivebrane give rise naively to (pseudo-)moduli, corresponding to
their motion in the $w$ plane, in which both kinds of fivebranes are extended. In the brane
description it is clear that these moduli are absent due to the attraction of the $D3$-branes
to the $NS$-brane \GiveonFK. Thus, the supersymmetry breaking vacuum of figure 4c is stable.

In four dimensions, the analog of the brane attraction in weakly coupled magnetic SQCD is the
one-loop potential for the pseudo-moduli computed in \IntriligatorDD. We expect something similar
to happen in the three dimensional case, but have not computed the potential for the pseudo-moduli
directly.

\newsec{Discussion}

In this note we proposed that $N=2$ supersymmetric level $k$ $U(N_c)$ Chern-Simons theory
with $N_f$ fundamental chiral superfields $Q^i$, $\tilde Q_i$ has a dual description, in
which the gauge group is replaced by $U(N_f+k-N_c)$, and the chiral superfields are
fundamentals $q_i$, $\tilde q^i$ as well as singlets $M^i_j$, coupled via the superpotential
\magsup. This duality exchanges regions with strong and weak CS coupling; in this sense,
it is a strong-weak coupling duality.

We presented the duality in terms of brane configurations in type IIB string theory, but it is
a property of CS theory. The brane description provides a convenient geometric language in terms
of which one can study the moduli spaces and deformations, both classically and quantum
mechanically, but the whole discussion could be repeated in field theory language.

A generalization of Seiberg duality to three dimensional $N=2$ supersymmetric gauge theory
was previously proposed in \refs{\KarchUX,\AharonyGP} (and further discussed from the brane
perspective in \GiveonSR). In these works the kinetic term of the gauge field had the standard
Yang-Mills form, and the CS term was absent. This leads to some differences with our analysis.

First, because the gauge coupling is dimensionful in three dimensions, in
\refs{\KarchUX,\AharonyGP} both the electric and the magnetic theories are strongly coupled
in the infrared. Thus, the dualities of \refs{\KarchUX,\AharonyGP} are strong-strong coupling
ones. Second, since the mass of the gauge field provided by the CS term is absent, there are
additional chiral superfields, associated with the vector superfield along the Coulomb branch
of the theory, which are difficult to define microscopically.

At the same time, the dualities of \refs{\KarchUX,\AharonyGP} are closely related to the one
described here. This is clear from the brane description we used. Indeed, before performing the
deformation of figure 1 for $k$ $D5$-branes on the electric and magnetic sides, the infrared
limits of the electric and magnetic brane configurations are precisely those of \AharonyGP.
In other words, assuming the dualities of \refs{\KarchUX,\AharonyGP}, our results can
be derived by turning on real masses for some of the flavors.

{}From this point of view, our main point is that turning on these real masses, eliminates
both of the problematic features of the dualities of \refs{\KarchUX,\AharonyGP}. By giving
a mass to the gauge field, it eliminates the Coulomb branch and the associated degrees of
freedom, and by replacing the Yang-Mills kinetic term with the CS one, it opens the possibility
of having a strong-weak coupling duality.

There are many questions along the lines of our discussion that require further work.
For example, in four dimensions, $N=1$ supersymmetric $SU(N_c)$ SYM theory with an adjoint
chiral superfield $X$ and $N_f$ fundamentals $Q^i$, $\tilde Q_i$, exhibits a generalization
of Seiberg duality when we turn on a polynomial superpotential for $X$, $W={\rm Tr} X^{p+1}$,
\refs{\KutasovVE\KutasovNP-\KutasovSS}.
This is related to the fact that in the theory with vanishing superpotential, the
dimension of the chiral operators ${\rm Tr} X^n$ can be made arbitrarily small
\refs{\IntriligatorJJ\KutasovIY\IntriligatorMI-\KutasovUX}. Some of the arguments
for the duality of \refs{\KutasovVE\KutasovNP-\KutasovSS} apply in three dimensions as well,
and it would be interesting to see whether there is a similar duality in this case.

There are of course many other known examples of Seiberg duality in four dimensions,
with or without string theory realizations, and it might be interesting to reexamine them in
the present context. More generally, Seiberg duality has many applications in  field and
string theory, some of which might be relevant in three dimensions as well.

Another interesting question of a more general nature is which combination of the $U(1)$
symmetries of an $N=2$ CS theory is the $U(1)_R$ that enters the superconformal multiplet
and determines the scaling dimensions of chiral operators. In four dimensions the answer
to this is given by a combination of considerations based on the NSVZ $\beta$-function,
$a$-maximization and Seiberg duality \refs{\IntriligatorJJ\KutasovIY\IntriligatorMI-\KutasovUX}.
In three dimensions, we have Seiberg duality, but the analog of NSVZ and $a$-maximization
is not available at present.

Finally, we commented briefly in section 4 on supersymmetry breaking in $N=2$ CS theory.
It is believed that many such theories have $AdS_4$ gravity duals \AharonyUG. It would
be interesting to understand the relation between spontaneous supersymmetry breaking in
the CS theory and its gravitational dual. This may help develop a holographic
understanding of four dimensional de Sitter vacua of the sort studied in \KachruAW.

\medskip
\noindent{\bf Note added:} After this work was completed, we received \AharonyGK,
where related issues were considered in the context of fractional $M2$-brane dynamics.

\bigskip
\noindent{\bf Acknowledgements:}
We thank O. Aharony for comments on the manuscript. 
This work was supported in part by the BSF -- American-Israel
Bi-National Science Foundation. AG is supported in part by a
center of excellence supported by the Israel Science Foundation
(grant number 1468/06), EU grant MRTN-CT-2004-512194, DIP
grant H.52, and the Einstein Center at the Hebrew University.
DK is supported in part by DOE grant DE-FG02-90ER40560 and the
National Science Foundation under Grant 0529954. DK thanks the
organizers of the Summer School on Particles, Fields and Strings
for hospitality during the conclusion of this work.

\listrefs
\end